\documentclass[fleqn,twoside,twocolumn,nofootinbib,showkeys,a4paper]{revtex4} 
\usepackage[nocpr]{ujp} 

\begin{document}
\title[Investigation of the Bosonic Spectrum]
{INVESTIGATION OF THE BOSONIC\\ SPECTRUM OF TWO-DIMENSIONAL
OPTICAL\\
GRAPHENE-TYPE LATTICES. NORMAL PHASE}
\author{I.V. Stasyuk}
\affiliation{Institute for Condensed Matter Physics, Nat. Acad. of
Sci. of Ukraine}
\address{1, Svientsitskii Str., Lviv 79011, Ukraine}
\author{I.R. Dulepa}%
\affiliation{Institute for Condensed Matter Physics, Nat. Acad. of
Sci. of Ukraine}%
\address{1, Svientsitskii Str., Lviv 79011, Ukraine}%
\author{O.V. Velychko}
\affiliation{Institute for Condensed Matter Physics, Nat. Acad. of
Sci. of Ukraine}
\address{1, Svientsitskii Str., Lviv 79011, Ukraine}
\udk{???} \pacs{37.10.Jk, 67.85.-d} \razd{\secvii}

\autorcol{I.V.\hspace*{0.7mm}Stasyuk, I.R.\hspace*{0.7mm}Dulepa,
O.V.\hspace*{0.7mm}Velychko}

\setcounter{page}{888}%

\begin{abstract}
The band spectrum of bosonic atoms in two-dimensional honeycomb
optical lattices with the graphene-type structure has been
studied.\,\,The dispersion laws in the bands and the one-particle
spectral densities are calculated for the normal phase in the random
phase approximation.\,\,The temperature-dependent gapless spectrum
with Dirac points located at the Brillouin zone boundary is obtained
for the lattice with energetically equivalent sites, with the
corresponding chemical potential lying outside the allowed energy
band.\,\,Different on-site energies in the sublattices are shown to
induce the appearance of a gap in the spectrum, so that the chemical
potential can be located between the subbands, which gives rise to a
substantial reconstruction of the band spectrum.\,\,The frequency
dependences of the one-particle spectral density for both
sublattices are determined as functions of the chemical potential
level, the spectral gap magnitude, and the temperature.
\end{abstract}
\keywords{optical lattice, {honeycomb} lattice, phase transition,
spectral density, hard-core bosons, Dirac points.} \maketitle

\section{Introduction}

Within the last decade, the considerable attention has been focused
on the research and the description of the phenomena occurring at
very low temperatures in subsystems of atoms that are located in the
so-called optical lattices. Such lattices are created under
laboratory conditions, using the interference of counter-propagating
coherent laser beams \cite{Greiner02a,Greiner02b}.\,\,The
electromagnetic field that arises in this case is periodic in space,
with its period being determined by the length of light waves  and
the relative angle between the beams.\,\,As a result, the potential
that acts on particles (atoms) in this field is also
periodic.\,\,Atoms in the optical lattice compose a perfect
quantum-mechanical system, almost all parameters of which can be
controlled.\,\,This fact makes it possible to study phenomena which
are hard to be observed in ordinary crystals.\,\,Depending on the
number and the orientation of interfering beams, one-, two-, and
three-dimensional lattices with various symmetries and structures
can be created \cite{Bloch05}.

Two important directions of modern quantum physics were combined to
research and to describe the behavior of ultracold Bose atoms in
two-dimensional optical lattices with the honeycomb structure.\,\,On
the one hand, in optical lattices, phase transitions associated with
the Bose condensation in the bosonic subsystem take place, and new
phases of specific types can also emerge.\,\,An additional interest
in such objects is related to the fact that a number of phenomena in
the physics of condensed state and systems with strong particle
correlations can be reproduced by analyzing the behavior of atoms
arranged in optical lattices.\,\,On the other hand, a
two-dimensional hexagonal carbon structure known as graphene became
the object of a special attention recently.\,\,It has the unique
physical properties resulting from the so-called Dirac energy
spectrum of conduction electrons (a linear dispersion law in a
vicinity of $K$-points in the Brillouin zone).\,\,Therefore, the
study of the thermodynamics and the energy spectrum of Bose atoms,
as well as Fermi ones, in optical lattices of the graphene type
attracts a considerable attention.\,\,The corresponding important
problems include, in particular, the research of how the mentioned
feature in the energy spectrum affects the scenario of phase
transitions in the system of ultra-cold atoms.\,\,The inverse
problem concerning a modification of the spectrum structure at the
transitions from one phase to the others is also of interest.

Quantum states in the system of bosonic atoms and a transition into
the phase with the Bose condensate (the so-called superfluid (SF)
phase) in an optical lattice of the graphene type were observed in
work \cite{a1}.\,\,The revealed regions of existence of various
phases (in the case concerned, these were the Mott insulator and the
SF phases) were in a qualitative agreement with the phase diagrams
calculated in the mean-field approximation.\,\,The specification of
phase region boundaries by making allowance for site-to-site
correlations with the help of the cluster generalization of the
Gutzwiller scheme was carried out later \cite{a2}.\,\,The attention
was also paid to honeycomb lattices; here, in contrast to graphene,
the states localized at the optical lattice sites are energetically
nonequivalent if those sites belong to different sublattices, $A$
and $B$.\,\,The cases of different on-site repulsion energies
($U_{A}\neq
U_{B}$) \cite{a3} and different potential well depths ($\varepsilon_{A}%
\neq\varepsilon_{B}$) \cite{a4,a5} were examined.\,\,In the latter
case, it was
taken into consideration that, besides $s$-states of atoms, the excited $p_{x,y}%
$-states of atoms localized in deeper wells can also participate in
the particle transfer and condensate formation processes.\,\,This
circumstance made it possible to study mechanisms governing the
formation of the so-called orbital (multiorbital) superfluid phase.

The features in the energy spectrum of bosons in optical lattices
with the graphene-type structure were considered in a few
works.\,\,In work \cite{a6}, changes in the arrangement of Dirac
points and the spectrum topology under the influence of the
interaction between particles were considered, and the weak coupling
approximation (in the framework of Bogolyubov's approach) was
applied.\,\,In works \cite{a5,a7}, the issues concerning the
displacement and the possible disappearance of Dirac points as a
result of the anisotropic ($t_{ij}\neq t_{ij^{\prime}}$) variations
of parameters for the particle transfer between the neighbor lattice
sites (such a variation can be stimulated by a mechanical shaking
\cite{a7}) were analyzed; however, a more complete analysis of the
spectrum and its reconstruction at transitions from one phases to
the others was not carri\-ed~out.

The theoretical description of the condensation of Bose particles in
optical lattices in general and, in particular, in lattices with the
graphene-type structure is mainly carried out on the basis of the
Bose--Hubbard model \cite{Fisher89,Jaksch98} and in its limiting
($U\rightarrow\infty$) case, the hard-core boson model
\cite{Whitlock63}.\,\,This model adequately describes the
thermodynamics and the energy spectrum of a bosonic system at low
population levels ($0\leq n\leq1$).\,\,Being applied (in the
simplest formulation) to honeycomb lattices, it enables one to find
the boundaries of the regions, where the main phases exist: Mott
insulator (MI), superfluid (SF) and modulated (CDW) phases; the
latter exists if the sublattices are nonequivalent.\,\,The extension
of the hard-core boson model by allowing the particle hopping
$t_{ij}$, besides the nearest, to farther lattice sites, revealed
the existence of new phases.\,\,As was shown in work \cite{a8}, a
large radius of the function $t_{ij}$ gives rise to the appearance
of a specific phase in the graphene-type lattice, the so-called Bose
metal.

The hard-core boson model is known already since the 1950s.\,\,Its
first application was associated with the liquid helium theory in
the framework of the lattice model \cite{a9}.\,\,The model was also
used in the theory of Josephson contact systems \cite{a10} and the
theory of high-temperature superconductivity (in the local-pair
approach) \cite{a11}.\,\,It was also made a basis for the
calculations of ionic conductivity in crystals \cite{a12}.\,\,During
last years, besides the description of the systems of ultracold Bose
particles in optical lattices, the model was also applied to study
the physical processes associated with ionic intercalation and
adsorption of quantum particles on a metal surface \cite{a13,a14}.

This work continues our theoretical researches
\cite{a15,a16,a17,a18} dealing with the energy spectrum and the
spectral characteristics of a quantum lattice Bose gas, and, in
particular, the hard-core boson model.\,\,In the framework of the
pseudospin approach, by applying the fermionization procedure in the
one-dimensional case \cite{a15} and the random phase approximation
in the more general three-dimensional one \cite{a16}, modifications
in the one-particle spectral densities at the transition from the
non-ordered (NO) state into the ordered one, in which $\langle
S^{x}\rangle=\langle b^{+}\rangle=\langle b\rangle\neq0$ and which
is an analog of the phase with the lattice Bose condensate (the SF
phase), were studied.\,\,The spectral densities and their frequency
dependences obtained in work \cite{a16} qualitatively agree with the
corresponding frequency dependences calculated on the basis of the
fermionization model and using the method of exact diagonalization
at one-dimensional clusters \cite{a17}.

Our present research aimed at studying the spectral characteristics
of a one-particle spectrum in the hard-core boson model in the case
of a plane honeycomb (of the graphene type) lattice with
energetically nonequivalent sites.\,\,A similar problem for a
three-dimensional lattice with a model density of states for the
nonperturbed one-particle spectrum was considered in work
\cite{a18}, where some general regularities in the structure of the
hard-core boson band spectrum were elucidated.\,\,The graphene-type
lattice, however, introduces its specificity into the spectrum
structure, and this issue had to be analyzed.\,\,We applied an
approach expounded in works \cite{a16,a18}.\,\,It is based on the
pseudo-spin formalism and the application of Green's function
technique while calculating the spectral densities.\,\,At the first
stage of calculations, the results of which are the topic of this
paper, we found a band structure and one-particle spectral densities
for the non-ordered (normal) phase and studied their dependences on
the location level of the chemical potential of Bose particles with
respect to the band spectrum, the difference between on-site
energies, $\delta=(\varepsilon_{A}-\varepsilon _{B})/2$, and the
temperature.

\section{Model}

In the general case, the Hamiltonian of the quantum lattice gas is given by
the expression%
\begin{equation}
H=-\sum_{i,j}t_{ij}b_{i}^{+}b_{j}+\sum_{i}(\varepsilon_{\alpha}-\mu)n_{i},
\end{equation}
where $t_{ij}$ is the transfer integral, $\varepsilon_{\alpha}$ are
the on-site energies ($\alpha=A$ or $B$ is the sublattice index),
$\mu$ is the chemical potential, $b_{i}^{+}\,(b_{i})$ is the
operator of particle creation (annihilation), and $n_{i}$ is the
number of particles at the $i$-th site.\,\,The site-to-site
interaction energy is neglected.

In the case of optical lattices with deep potential wells, the
energy of the on-site repulsion of Bose atoms is high, so that the
hard-core boson model, in which the site population number is
restricted ($n_{i}=0$ or 1), is a good approximation.\,\,Such bosons
are described by the Pauli operators with the commutation relations
\begin{equation}
 [b_i^+,b_j^+]\!=\![b_i,b_j]\!=\![b_i^+,b_j]\!=\!0,~ i \!\neq\! j;~ \{b_i,b_i^+\}
    \!=\!
    1.\!\!\!\!
\end{equation}
The model becomes equivalent to the problem with the pseudospin $S={\downarrow
}$ ($S={\uparrow}$) after the transformation%
\begin{equation}
 b_i^+=S_i^-,~ b_i=S_i^+,~ b_i^+ b_i = n_i= \frac{1}{2} - S_i^z.
\end{equation}
In the spin representation, the Hamiltonian looks like%
\begin{equation}
H=-\sum_{i,j}t_{ij}S_{i}^{-}S_{j}^{+}-\sum_{i}h_{\alpha}S_{i}^{z}%
+\mathrm{const},
\end{equation}
where\vspace*{-3mm}
\begin{equation}
 h_\alpha = (\varepsilon_\alpha - \mu),\quad
    \mathrm{const} =\sum_{\alpha=A,B} (\varepsilon_\alpha -
            \mu)\frac{N}{2}.
\end{equation}
Below, the constant term in the Hamiltonian is omitted.\,\,The
summation over $i$ implies the summation over the cell index $n$ and
the sublattice index $\alpha$.

Taking the aforesaid into account, in the case of two sublattices
($\alpha=A,B$), we obtain the following expression for the Hamiltonian:%
\[H =-\sum_{nn'} J_{nn'}^{AB}(S_{nA}^x S_{n'B}^x + S_{nA}^y S_{n'B}^y)\, -
\]\vspace*{-5mm}
\[-\sum_{nn'} J_{nn'}^{BA}(S_{nB}^x S_{n'A}^x + S_{nB}^y S_{n'A}^y)\, -
\]\vspace*{-5mm}
\begin{equation}
-\,h_A \sum_n S_{nA}^z - h_B \sum_{n'} S_{n'B}^z .
\end{equation}
Making a rotation by a certain angle $\theta$ in the spin space,
\begin{align}
&
S_{n\alpha}^{z}=\sigma_{n\alpha}^{z}\cos\theta_{\alpha}+\sigma_{n\alpha
}^{x}\sin\theta_{\alpha}\nonumber\\[1mm] &
S_{n\alpha}^{x}=\sigma_{n\alpha}^{x}\cos\theta_{\alpha}-\sigma_{n\alpha
}^{z}\sin\theta_{\alpha},\\[1mm]
&  S_{n\alpha}^{y}=\sigma_{n\alpha}^{y},\nonumber
\end{align}
we obtain
\begin{align}
  &H = \notag\\
  &=\!-\!\sum_{nn'}\!\left[L_1^{AB}(n,n^\prime) \sigma_{nA}^x \sigma_{n'B}^x
  \!+\!
                  L_2^{AB}(n,n^\prime) \sigma_{nA}^z \sigma_{n'B}^z\right]+ \notag\\
     &+\sum_{nn'}\left[L_3^{AB}(n,n^\prime) \sigma_{nA}^x \sigma_{n'B}^z +
                  L_4^{AB}(n,n^\prime) \sigma_{nA}^z \sigma_{n'B}^x\right] - \notag\\
     &-\sum_{nn'}L_5^{AB}(n,n^\prime) \sigma_{nA}^y \sigma_{n'B}^y \,- \notag\\
     &-\sum_\alpha h_\alpha \sum_{n}(\sigma_{n\alpha}^z\cos\theta_\alpha +
       \sigma_{n\alpha}^x \sin\theta_\alpha),
\end{align}
where the notations
\begin{align}
&L_1^{AB}(n,n^\prime)=(J_{nn'}^{AB}+J_{n'n}^{BA})\cos\theta_A \cos
                   \theta_B,\notag\\[1mm]
&L_2^{AB}(n,n^\prime)=(J_{nn'}^{AB}+J_{n'n}^{BA})\sin\theta_A \sin
                   \theta_B,\notag\\[1mm]
&L_3^{AB}(n,n^\prime)=(J_{nn'}^{AB}+J_{n'n}^{BA})\cos\theta_A \sin
                   \theta_B,\\[1mm]
&L_4^{AB}(n,n^\prime)=(J_{nn'}^{AB}+J_{n'n}^{BA})\sin\theta_A \cos
                   \theta_B,\notag\\[1mm]
&L_5^{AB}(n,n^\prime)=J_{nn'}^{AB}+J_{n'n}^{BA}.\notag
\end{align}
were introduced. Carrying out the Fourier transformation, we change to the
wavevectors,%
\begin{equation}
\begin{array}{l}
      \displaystyle \sum_{n'}(J_{nn'}^{AB}+J_{n'n}^{BA})\mathrm{e}^{i\mathbf{q}(\mathbf{R}_{nA}-\mathbf{R}_{n'B})}\,
       =\\[2mm] =\,J^{AB}(\mathbf{q})
       \equiv J(\mathbf{q}),\\[2mm]
     \displaystyle   \sum_{n'}(J_{nn'}^{BA}+J_{n'n}^{AB})\mathrm{e}^{i\mathbf{q}(\mathbf{R}_{nB}-\mathbf{R}_{n'A})}\,
       =\\[2mm]=\, J^{BA}(\mathbf{q})
       \equiv J(-\mathbf{q}).
\end{array}\!\!\!\!\!\!\!\!\!\!\!\!\!\!\!\!\!\!\!\!
\end{equation}
Then
\begin{align}
   &L_1^{AB}(\mathbf{q})=J(\mathbf{q})\cos\theta_A \cos
                \theta_B, \notag\\
   &L_2^{AB}(\mathbf{q})=J(\mathbf{q})\sin\theta_A \sin
                \theta_B, \notag\\
   &L_3^{AB}(\mathbf{q})=J(\mathbf{q})\cos\theta_A \sin
                \theta_B, \\
   &L_4^{AB}(\mathbf{q})=J(\mathbf{q})\sin\theta_A \cos
                \theta_B,\notag\\
   &L_5^{AB}(\mathbf{q})=J(\mathbf{q}).\notag
\end{align}
Taking into account that the environments of sites belonging to
different sublattices are equivalent, we may write
$J^{AB}(0)=J^{BA}(0)\equiv J(0)=3t$, where $t$ is the doubled
transfer integral between neighbor lattice sites (see
Appendix\,\,A).\,\,In the Hamiltonian, we single out a part that
corresponds to the
mean-field (MF) approximation,%
\begin{equation}
\begin{array}{l}
\sigma_{nA}^\nu \sigma_{n' B}^{\nu'}
   \to \langle \sigma_A^\nu \rangle \sigma_{n'
  B}^{\nu'} + \langle \sigma_B^{\nu'} \rangle \sigma_{n
  A}^{\nu} - \langle \sigma_{A}^\nu \rangle \langle
  \sigma_{B}^{\nu'} \rangle, \\[2mm]
\nu,\nu'=x,y,z, \langle \sigma_{\alpha}^{x}
  \rangle=\langle \sigma_{\alpha}^{y} \rangle=0.
  \end{array}
\end{equation}
As a result, the mean-field Hamiltonian reads%
\begin{equation}
H_{\text{MF}}=-\sum_{n\alpha}E_{\alpha}\sigma_{n\alpha}^{z}.
\end{equation}
The corresponding eigenvalues and the rotation angles $\theta_{\alpha}$ are
determined from the system of equations%
\begin{equation}
\begin{array}{l}
\label{eq001} E_\alpha =
            (\varepsilon_\alpha - \mu) \cos \theta_\alpha -
            J (0)\langle S_\beta^x\rangle \sin \theta_\alpha,
            \\[2mm]
(\varepsilon_\alpha - \mu)  \sin \theta_\alpha + J (0)\langle
            S_\beta^x\rangle \cos \theta_\alpha =0,
\end{array}
\end{equation}
where%
\begin{equation}
\begin{array}{l}
\label{eq002}\langle S_\alpha^x\rangle = -\langle
\sigma_\alpha^z\rangle \sin
  \theta_\alpha,\,\langle S_\alpha^z\rangle = \langle
  \sigma_\alpha^z\rangle \cos \theta_\alpha, \\[2mm]
\langle \sigma_\alpha^z\rangle = \frac12 \tanh\frac{\beta
  E_\alpha}{2},\ \alpha,\beta = A,B, \, \alpha \neq \beta.
\end{array}
\end{equation}

In the non-ordered phase (for the system of bosons, this is the
so-called normal phase), $\theta_{\alpha}=0$, $\langle
S_{\alpha}^{x}\rangle=0$, $\langle
S_{\alpha}^{z}\rangle=\langle\sigma_{\alpha}^{z}\rangle$, and
$E_{\alpha}=\varepsilon_{\alpha}$.\,\,The solution
$\theta_{\alpha}\neq0$ describes the \textquotedblleft
ordered\textquotedblright\ phase (the phase
with the condensate of hard-core bosons), for which $\langle S_{\alpha}%
^{x}\rangle\equiv$ $\equiv\langle b_{\alpha}\rangle\neq0$ is the
order parameter.\,\,The system of equations (14), together with
formulas (\ref{eq002}), determines the behavior of the order
parameter and the average $\langle S_{\alpha}^{z}\rangle$,
i.e.\,\,$\langle n_{\alpha}\rangle$, as the temperature in the
ordered phase varies.\,\,The temperature-induced variation of the
order parameter $\langle S^{x}\rangle$ in the case where the crystal
is not separated into sublattices \cite{a2}, for the given on-site
energy, and in the mean-field approximation is the same as in the
Ising model with the transverse field acting on the spin (the  role
of the field in this work is played by the quantity
$h_{\alpha}=\varepsilon_{\alpha}-\mu$).\,\,In the further
calculations, we will study the bosonic band spectrum in the
non-ordered (NO) phase at a fixed temperature and its dependence on
the fields $h_{\alpha}$ at various distances from the curves on the
phase diagrams (see work \cite{a18}) that correspond to the
transitions into the phase with a Bose condensate (the SF phase).

\section{Green's Functions\\ and the Energy Spectrum of the Model}

The one-particle energy spectrum can be calculated using the Green's
function method and the random phase approximation.\,\,The
one-particle Green's function on the operators $\langle\langle
b_{l\alpha}|b_{n\beta}^{+}\rangle\rangle$ equals
Green's function on the pseudospin operators $\langle\langle S_{l\alpha}%
^{+}|S_{n\beta}^{-}\rangle\rangle\equiv$\linebreak $\equiv
G_{l\alpha,n\beta}^{+-}$
\mbox{\cite{a16,a18}}:%
\[
  \langle\langle S_{l\alpha}^+|S_{n\beta}^-\rangle\rangle =
  \langle\langle S_{l\alpha}^x|S_{n\beta}^x\rangle\rangle - i
  \langle\langle S_{l\alpha}^x|S_{n\beta}^y\rangle\rangle\, + \]\vspace*{-7mm}
\begin{equation}
  +\, i\langle\langle S_{l\alpha}^y|S_{n\beta}^x\rangle\rangle +
  \langle\langle S_{l\alpha}^y|S_{n\beta}^y\rangle\rangle.
\end{equation}
In the NO-phase ($\cos\theta_{\alpha}=1,\,\sin\theta_{\alpha}=0$),%
\[
G_{l\alpha,n\beta}^{+-} = \langle\langle
\sigma_{l\alpha}^x|\sigma_{n\beta}^x\rangle\rangle - i
\langle\langle \sigma_{l\alpha}^x|\sigma_{n\beta}^y\rangle\rangle
\,+
\]\vspace*{-7mm}
  \begin{equation}
+\, i \langle\langle
\sigma_{l\alpha}^y|\sigma_{n\beta}^x\rangle\rangle+ \langle\langle
\sigma_{l\alpha}^y|\sigma_{n\beta}^y\rangle\rangle.
\end{equation}
The equation of motion for Green's functions in the pseudospin component
representation looks like%
\begin{equation}
\begin{array}{l}
\displaystyle\hbar \omega \langle\langle
\sigma_{l\alpha}^{\nu}|\sigma_{n\beta}^{\nu'}\rangle\rangle \!=\!
\frac{\hbar}{2\pi}
\langle[\sigma_{l\alpha}^{\nu},\sigma_{n\beta}^{\nu'}]\rangle \!+\!
\langle\langle[\sigma_{l\alpha}^{\nu},H]|\sigma_{n\beta}^{\nu'}\rangle\rangle,
\\[5mm]
\nu,\nu'=x,y.
\end{array}\!\!\!\!\!\!\!\!\!\!\!\!\!\!\!\!\!\!\!\!
\end{equation}
Let us perform the decoupling of Green's function of the higher
order, which corresponds to the random phase approximation.\,\,At
this decoupling, $[\sigma_{l\alpha}^{z},H]\rightarrow$
$\rightarrow0$ and, as a result, $[\sigma_{l\alpha
}^{z},H]\rightarrow0$.\,\,For Green's functions with transverse
pseudospin
components, we obtain the system of equations%
\begin{equation}
\begin{array}{l}
\label{seq} \displaystyle\hbar \omega \langle\langle
  \sigma_{l\alpha}^x|\sigma_{n\alpha}^x\rangle\rangle  = iE_\alpha
  \langle\langle \sigma_{l\alpha}^y|\sigma_{n\alpha}^x\rangle\rangle\, -
  \\[3mm]
 \displaystyle-\, i \langle \sigma_\alpha^z\rangle \sum_{n'} L_5^{\alpha\beta}
  \langle\langle
  \sigma_{n'\beta}^y|\sigma_{n\alpha}^x\rangle\rangle,\\[1mm]
\displaystyle\hbar \omega \langle\langle
  \sigma_{l\alpha}^y|\sigma_{n\alpha}^x\rangle\rangle  = -i \frac{\hbar}{2\pi}
  \langle \sigma_\alpha^z\rangle \delta_{ln}
  - i E_\alpha
  \langle\langle \sigma_{l\alpha}^x|\sigma_{n\alpha}^x\rangle\rangle\, +
  \\[3mm]
\displaystyle +\, i\langle \sigma_\alpha^z\rangle \sum_{n'} L_1^{\alpha
\beta}
  \langle\langle
  \sigma_{n'\beta}^x|\sigma_{n\alpha}^x\rangle\rangle, \\[3mm]
\displaystyle\hbar \omega \langle\langle
  \sigma_{l\beta}^x|\sigma_{n\alpha}^x\rangle\rangle  = iE_\beta
  \langle\langle \sigma_{l\beta}^y|\sigma_{n\alpha}^x\rangle\rangle \,- \\[3mm]
\displaystyle-\, i \langle \sigma_\beta^z\rangle \sum_{n'}
L_5^{\beta\alpha}
  \langle\langle
  \sigma_{n'\alpha}^y|\sigma_{n\alpha}^x\rangle\rangle,  \\[3mm]
\displaystyle\hbar \omega \langle\langle
  \sigma_{l\beta}^y|\sigma_{n\alpha}^x\rangle\rangle  = -iE_\beta
  \langle\langle \sigma_{l\beta}^x|\sigma_{n\alpha}^x\rangle\rangle\, + \\[3mm]
 \displaystyle+\, i \langle \sigma_\beta^z\rangle \sum_{n'} L_1^{\beta\alpha}
  \langle\langle \sigma_{n'\alpha}^x|\sigma_{n\alpha}^x\rangle\rangle .
\end{array}\!\!\!\!\!\!\!\!\!\!\!\!\!\!\!\!\!\!\!\!
\end{equation}
Hereafter, $\alpha\neq\beta$.\,\,The system of equations for the
functions
$\langle\langle\sigma_{\ldots}^{y}|\sigma_{n\alpha}^{x}\rangle\rangle$
has a
similar form.\,\,After the Fourier transformation to the wave vectors,%
\begin{equation}
G_{\alpha\beta}^{\nu\nu^{\prime}}({\mathbf{q}})\equiv\sum_{l-n}\langle
\langle\sigma_{l\alpha}^{\nu}|\sigma_{n\beta}^{\nu^{\prime}}\rangle
\rangle\mathrm{e}^{i{\mathbf{q}}({\mathbf{R}}_{l\alpha}-{\mathbf{R}}_{n\beta
})},
\end{equation}
where
$L_{1}^{AB}({\mathbf{q}})=L_{5}^{AB}({\mathbf{q}})=J({\mathbf{q}})$
and $L_{1}^{BA}(\mathbf{q})=$\linebreak
$=L_{5}^{BA}(\mathbf{q})=J(-\mathbf{q})$, the system of equations
(\ref{seq}) reads
\begin{equation}
\begin{array}{l}
  \hbar \omega G_{\alpha\alpha}^{xx}({\mathbf q})= i E_\alpha
  G_{\alpha\alpha}^{yx}({\mathbf q})-i J({\mathbf q})\langle
  \sigma_\alpha^z
  \rangle G_{\beta\alpha}^{yx}({\mathbf q}),  \\[2mm]
 \displaystyle \hbar \omega G_{\alpha\alpha}^{yx}({\mathbf q}) = -i
  \frac{\hbar}{2 \pi} \langle \sigma_\alpha^z \rangle -i E_\alpha
  G_{\alpha\alpha}^{xx}({\mathbf q})\, +  \\[4mm]
   +\, i J({\mathbf q})\langle \sigma_\alpha^z
  \rangle G_{\beta\alpha}^{xx}({\mathbf q}), \\[3mm]
  \hbar \omega G_{\beta\alpha}^{xx}({\mathbf q})= i E_\beta
  G_{\beta\alpha}^{yx}({\mathbf q}) - i J(-{\mathbf q})\langle
  \sigma_\beta^z \rangle G_{\alpha\alpha}^{yx}({\mathbf q}),  \\[3mm]
  \hbar \omega G_{\beta\alpha}^{yx}({\mathbf q})  = -i E_\beta
  G_{\beta\alpha}^{xx}({\mathbf q}) + i J(-{\mathbf q})\langle
  \sigma_\beta^z \rangle G_{\alpha\alpha}^{xx}({\mathbf q}) .
\end{array}\!\!\!\!\!\!\!\!\!\!\!\!\!\!\!\!\!\!\!\!
\end{equation}
The system of equations for Green's functions $G_{\alpha\alpha}^{\nu
y}({\mathbf{q}})$ and $G_{\beta\alpha}^{\nu y}({\mathbf{q}})$ has a
similar form (with the substitution $J({\mathbf{q}})\rightarrow
J(-{\mathbf{q}})$ at proper places).\,\,The sought Green's function
is
\begin{equation}
\label{eq003}
    G_\alpha^{+-}({\mathbf q})=\langle\langle
    b_\alpha|b_\alpha^+\rangle\rangle_{\mathbf q}=
    G_\alpha^{+x}({\mathbf q})-iG_\alpha^{+y}({\mathbf q}).
\end{equation}
Here,%
\begin{equation}
    G_\alpha^{+x}({\mathbf q})= \langle\langle
    \sigma_\alpha^x|\sigma_\alpha^x\rangle\rangle_{\mathbf q} + i\langle\langle
    \sigma_\alpha^y|\sigma_\alpha^x\rangle\rangle_{\mathbf q},
    \end{equation}\vspace*{-7mm}
   \begin{equation}
G_\alpha^{+y}({\mathbf q})= \langle\langle
    \sigma_\alpha^x|\sigma_\alpha^y\rangle\rangle_{\mathbf q} + i\langle\langle
    \sigma_\alpha^y|\sigma_\alpha^y\rangle\rangle_{\mathbf q}.
\end{equation}

The equations given above have the following solutions:
\[G_{\alpha\alpha}^{\pm x}(\omega, {\mathbf q}) =  \pm  \frac{\hbar}{2 \pi} \langle \sigma_\alpha^z \rangle\, \times\]\vspace*{-7mm}
\begin{equation}
\times\,\frac{\hbar\omega \mp E_\beta}{(\hbar \omega -
E_\alpha)(\hbar \omega - E_\beta) - \langle \sigma_\alpha^z \rangle
\langle \sigma_\beta^z \rangle |J({\mathbf q})|^2},
\end{equation}\vspace*{-7mm}
\[
G_{\alpha\alpha}^{\pm y}(\omega, {\mathbf q}) = i \frac{\hbar}{2
\pi} \langle \sigma_\alpha^z \rangle\, \times\]\vspace*{-7mm}
\begin{equation}
\times\,\frac{\hbar\omega \mp
 E_\beta}{(\hbar \omega - E_\alpha)(\hbar \omega - E_\beta) -
 \langle \sigma_\alpha^z \rangle \langle \sigma_\beta^z \rangle
 |J({\mathbf q})|^2}.
\end{equation}
The ultimate expressions for one-particle Green's functions are
\begin{equation}
G_{\beta\alpha}^{+-}(\omega, {\mathbf q}) =
 -\frac{\langle \sigma_\beta^z \rangle J({\mathbf q})}{\hbar \omega - E_{\beta}}
 G_{\alpha\alpha}^{+-}(\omega, {\mathbf q}),
\end{equation}\vspace*{-7mm}
\[
G_{\alpha\alpha}^{+-}(\omega, {\mathbf q}) =
\frac{\hbar}{\pi}\langle \sigma_\alpha^z
\rangle\,\times\]\vspace*{-7mm}
\begin{equation}\vspace*{-7mm}
\label{eq004}
 \times \,  \frac{\hbar \omega -
 E_\beta}{(\hbar \omega -E_\alpha)(\hbar \omega - E_\beta) -
 \langle \sigma_\alpha^z \rangle \langle \sigma_\beta^z \rangle
 |J({\mathbf q})|^2}.
\end{equation}

In the normal phase, the spectrum of bosonic excitations determined from
Eq.\,(\ref{eq004}) looks like (see also work \cite{a18})%
\[
 \varepsilon_{1,\,2}({\mathbf q}) = \frac{h_A + h_B}{2}\,\pm\]\vspace*{-7mm}
\begin{equation}
 \pm\,   \frac12\sqrt{(h_A - h_B)^2 + 4\langle \sigma_A^z \rangle
      \langle \sigma_B^z \rangle |J({\mathbf q})|^2},
\end{equation}\vspace*{-7mm}
\[
  J({\mathbf q})=t \left( \!\mathrm{e}^{iq_ya} + 2 \mathrm{e}^{-iq_ya} \cos \left(\!\textstyle\frac{\sqrt{3}}{2}a q_x \!\right)\! \right)
\]
(see Appendix~A).\,\,Using the notations $h=\frac{h_{A}+h_{B}}{2}$
and$\,\delta =\frac{h_{A}-h_{B}}{2}$, the expression for the
spectrum can be written in the
form%
\begin{equation}
\varepsilon_{1,\,2}({\mathbf{q}})=h\pm\sqrt{\delta^{2}+\frac{1}{9}%
\langle\sigma_{A}^{z}\rangle\langle\sigma_{B}^{z}\rangle J^{2}(0)|\gamma
({\mathbf{q}})|^{2}}. \label{eq005}%
\end{equation}

The regions and the boundaries of existence for the normal (NO)
phase, as well as for the phase with the Bose condensate (the SF
phase), follow from the divergence condition for the function
$G_{\alpha\alpha}^{+-}$ at $\omega\rightarrow0$ and
$\mathbf{q}\rightarrow0$.\,\,The corresponding equation looks like
\begin{equation}
h^{2}-\delta^{2}=\langle\sigma_{A}^{z}\rangle\langle\sigma_{B}^{z}\rangle
J^{2}(0)\equiv\langle\sigma_{A}^{z}\rangle\langle\sigma_{B}^{z}\rangle9t^{2}.
\end{equation}
The relevant $(T,h)$ phase diagram in terms of $J(0)$ units is
plotted in Fig.~1.

\section{Spectrum of Bosonic Excitations. One-Particle Spectral Density of
States}

Let us determine the spectral density of bosonic excitations per one $\alpha
$-sublattice site ($\alpha=A,B$) for both sublattices as the imaginary part of
Green's function $\langle\langle b_{i\alpha}|b_{i\alpha}^{+}\rangle
\rangle_{\omega+i\varepsilon}$:%
\begin{equation}
\rho_{\alpha}(\omega)=-\frac{1}{N}\sum_{\mathbf{q}}\text{Im}\,\langle\langle
b_{\alpha}|b_{\alpha}^{+}\rangle\rangle_{\mathbf{q},\omega+i\varepsilon}.
\end{equation}
On the basis of Eq.~(\ref{eq004}), we obtain%
\[
\rho_\alpha (\omega) =
\frac{\langle\sigma_\alpha^z\rangle}{N}\sum_{\mathbf q} \bigg(\!
{C_1(\mathbf q) \delta \bigg(\!\omega - \frac{\varepsilon_1(\mathbf
q)}{\hbar}\!\bigg)}+\]\vspace*{-5mm}
\begin{equation}
+\, {C_2(\mathbf q) \delta \bigg(\!\omega -
\frac{\varepsilon_2(\mathbf q)}{\hbar}\!\bigg)}\!\bigg)\!,
\end{equation}
where the coefficients before $\delta$-functions equal%
\begin{equation}
C_{1,2}(\mathbf{q})=\frac{1}{2}\pm\frac{\delta_{\alpha}}{2\sqrt{\delta
^{2}+\frac{1}{9}\langle\sigma_{\alpha}^{z}\rangle\langle\sigma_{\beta}%
^{z}\rangle J^{2}(0)|\gamma({\mathbf{q}})|^{2}}}.
\end{equation}
Here, $\alpha\neq\beta$,%
\[
\delta_{\alpha}=\left\{\!\!
\begin{array}
[c]{ll}%
\delta, & \text{$\alpha=A$},\\
-\delta, & \text{$\alpha=B$},
\end{array}
\right.
\]
and $\varepsilon_{1}(\mathbf{q})$ and
$\,\varepsilon_{2}(\mathbf{q})$ are the branches of spectrum
(\ref{eq005}).\,\,This expression for the spectral density in the NO
phase formally coincides with that obtained in work \cite{a18} for
the case of a cubic lattice.

\begin{figure}[t]%
\vskip1mm
  \includegraphics[width=\column]{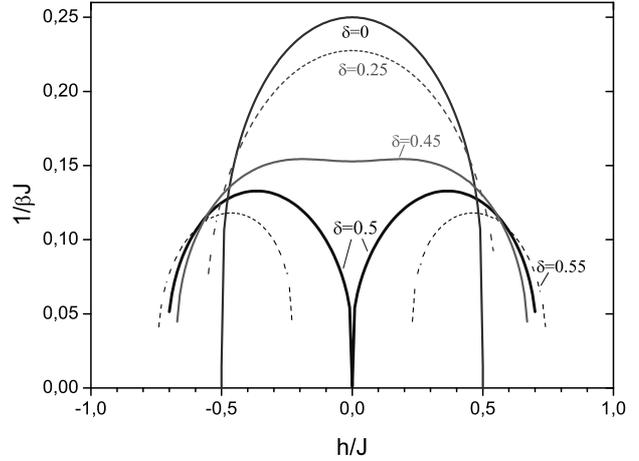}
  \vskip-3mm
\caption{Phase diagrams in the plane $(T,h)$ for various values
{$\delta=0$, 0.25, 0.45, 0.5, and 0.55} {\cite{a18}}.\,\,In this and
other figures, all energy quantities are reckoned in
$J(0)$-units}\vspace*{-2mm}
\end{figure}

The dependence of $\rho_{\alpha}(\omega,{\mathbf{q}})$ on the wave
vector is expressed through the dependence of $J({\mathbf{q}})$ on
$\mathbf{q}$. The summation over $\mathbf{q}$ is carried out within
the limits of the first Brillouin zone $\Omega$.\,\,In order to
calculate this sum, we change to the integral over the variable
$x\equiv$ $\equiv|\gamma_{\mathbf{q}}|^{2}$ and introduce
the function $\rho_{0}(x)$:\vspace*{-1mm}%
\begin{align}
&  \frac{1}{N}\sum_{{\mathbf{q}}\in\Omega}\Phi(|J({\mathbf{q}})|^{2})=\frac
{1}{N}\sum_{{\mathbf{q}}\in\Omega}\Phi(t^{2}|\gamma({\mathbf{q}}%
)|^{2})=\nonumber\\
&  =\int\mathrm{d}x\rho_{0}(x)\Phi(t^{2}x),\nonumber\\
&  \rho_{0}(x)=\frac{1}{N}\sum_{\mathbf{q}\in\Omega}\delta(x-|\gamma
({\mathbf{q}})|^{2}).
\end{align}\vspace*{-3mm}

\noindent The transition from the summation over $\mathbf{q}$ to
the integation within the first
Brillouin zone $\Omega$ is done according to the formula\vspace*{-1mm}%
\begin{equation}
\frac{1}{N}\sum_{{\mathbf q }\in \Omega} \left(...\right)=
\frac{S}{(2\pi)^2 N} \int\limits_\Omega \mathrm{d}q_x \mathrm{d}q_y
\left(...\right)\!,
\end{equation}\vspace*{-3mm}

\noindent where $S$ is the area of the so-called main crystal
region, and $N$ is the number of cells.\,\,The sense of the ratio
$S/N$ is the area of the elementary cell formed by the vectors
${\mathbf{a}_{1}}$ and ${\mathbf{a}_{2}}$ in the coordinate space:
$\left\vert {\mathbf{a}_{1}}\right\vert =\left\vert
{\mathbf{a}_{2}}\right\vert =a\sqrt{3}$, $\frac{S}{N}=\frac{3\sqrt{3}}{2}%
a^{2}$.

Let us consider the integration limits over $q_{x}$ and
$q_{y}$.\,\,From Fig.~2, one can see that, instead of the
integration over the region $\Omega$, it is possible to integrate
within the limits of the marked rectangle.\,\,Since the integrand is
an even function of the variables $q_{x}$ and $q_{y}$, for
the summation over ${\mathbf{q}}\in\Omega,$ we have\vspace*{-1mm}%
\begin{equation}
\frac{1}{N}\sum_{\mathbf q} \left(...\right)= \frac{3 \sqrt{3} a^2}{
(2 \pi)^2}  \int\limits_0^{\frac{2\pi}{\sqrt{3}a}} \mathrm{d}q_x
\int\limits_0^{\frac{2\pi}{3a}} \mathrm{d}q_y \left(...\right)\!.
\end{equation}\vspace*{-3mm}

\noindent In terms of the variables
$2\vartheta=\frac{\sqrt{3}}{2}q_{x}a$ and
$\varphi=\frac{3}{2}q_{y}a$, this formula looks like\vspace*{-1mm}%
\begin{equation}
 \frac{1}{N} \sum_{\mathbf q} (...) = \frac{2}{\pi^2}
\int\limits_0^\frac{\pi}{2} \mathrm{d}\vartheta \int\limits_0^\pi
\mathrm{d}\varphi (...).
\end{equation}\vspace*{-3mm}

\noindent
The final expression for $\rho_{0}(x)$ in the case concerned has the form\vspace*{-2mm}%
\[ \rho_0(x)=\frac{1}{\pi^2}\int\limits_0^\pi \mathrm{d}\vartheta \int\limits_0^\pi
\mathrm{d} \varphi\, \times \]\vspace*{-7mm}
\begin{equation}
\label{eq007} \times\, \delta(x-1-4\cos2\vartheta \cos \varphi -
4\cos^2 2\vartheta).
\end{equation}
Formula (\ref{eq007}) directly corresponds to the expression for the
distribution function over the squared energy, $g(\varepsilon^{2})$,
for noninteracting particles in the lattice with the graphene-type
structure \cite{a19,a20}, according to which $\rho_{0}(x)$ can be
expressed by means of
the complete elliptic integral of the first kind, $F(\frac{\pi}{2},m)$:%
\begin{equation}
\rho_{0}(x)=\frac{1}{\pi^{2}}\frac{1}{\sqrt{Z_{0}}}F\left(\!  \frac{\pi}%
{2},\sqrt{\frac{Z_{1}}{Z_{0}}}\!\right)\!  ,
\end{equation}

\begin{figure}%
\vskip1mm \includegraphics[width=5.5cm]{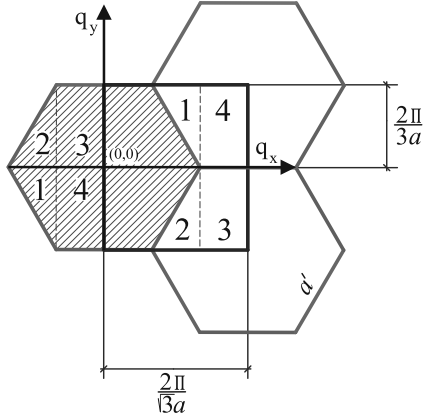}
\vskip-3mm
\caption{First Brillouin zone $\Omega$ in the reciprocal lattice
(identical figures mark translationally equivalent regions)}
\end{figure}

\noindent where\vspace*{-3mm}
\begin{displaymath}
      {Z_0} =   \left\{\!\!
                \begin{array}{ll}
                             (1+\sqrt{x} )^2 -\frac14 (x-1)^2, & x \leqslant 1,
                             \\[1mm]
                                                    4\sqrt {x}, & 1\leqslant  x \leqslant 9;
               \end{array}
               \right.
\end{displaymath}\vspace*{-6mm}
\begin{displaymath}
       {Z_1} =  \left\{\!\!
                \begin{array}{ll}
                             4\sqrt {x}, & x \leqslant 1, \\[1mm]
        (1+\sqrt{x})^2 -\frac14 (x-1)^2, & 1 \leqslant  x \leqslant 9.
                \end{array}
                \right.
\end{displaymath}
The obtained function can be used to calculate the spectral density,%
\[\rho_\alpha (\omega) = \langle\sigma_\alpha^z\rangle
\int\mathrm{d}x \rho_0(x)\bigg(\!{C_1(x) \delta \bigg(\!\omega -
 \frac{\varepsilon_1(\mathbf q)}{\hbar}\!\bigg)}\,+
\]\vspace*{-7mm}
\begin{equation}
{}+ {C_2(x) \delta \bigg(\!\omega - \frac{\varepsilon_2(\mathbf
q)}{\hbar}\!\bigg)}\!\bigg)\!,
\end{equation}\vspace*{-5mm}

\noindent where\vspace*{-3mm}
\begin{equation}
C_{1,2}(x)=\frac{1}{2}\!\left(\!1 \pm
\frac{\delta_\alpha}{\sqrt{\delta^2 + \langle\sigma_\alpha^z\rangle
\langle\sigma_\beta^z \rangle t^2 x}}\!\right)\!\!,
\end{equation}\vspace*{-7mm}
\begin{equation}
\langle\sigma_\alpha^z\rangle = \frac12 \tanh\frac{\beta
h_\alpha}{2}.
\end{equation}

For the $\delta$-functions in the expression for
$\rho_{\alpha}(\omega)$, we use the formula
$\delta(f(x))=\sum_{i}\frac{\delta(x-x_{i})}{|f^{\prime}(x_{i})|}$,
where $x_{i}$ are the roots of the equation $f(x)=0$.\,\,In our
case,
\[
x_{0}=\frac{(\hbar\omega-h)^{2}-\delta^{2}}{\langle\sigma_{\alpha}^{z}%
\rangle\langle\sigma_{\beta}^{z}\rangle t^{2}}%
\]
is a root for both $\delta$-functions, with the first one giving a
nonzero contribution at $\hbar\omega>h$, and the second one at
$\hbar\omega<h$.\,\,The corresponding derivative
\[
\lvert f^{\prime}(x_{1,2})\rvert=\frac{t^{2}}{\hbar}\left\vert \frac
{\langle\sigma_{\alpha}^{z}\rangle\langle\sigma_{\beta}^{z}\rangle}%
{2(\hbar\omega-h)}\right\vert.
\]

\noindent
After simplifications, we obtain%
\begin{equation}
\rho_{\alpha}(\hbar\omega)=\frac{\rho_{\alpha}(\omega)}{\hbar}=\frac
{\langle\sigma_{\alpha}^{z}\rangle}{t^{2}}\Big(\!\rho_{\alpha}^{(1)}%
(\omega)+\rho_{\alpha}^{(2)}(\omega)\!\Big)\!.
\end{equation}
Here, the spectral density per unit energy interval was
introduced:\vspace*{-3mm}
\begin{equation}
\label{eq006} \rho_\alpha^{(1,2)}(\omega) = \rho_0(x_0) \left\lvert
\frac{\hbar \omega - h}{\langle \sigma_\alpha^z \rangle \langle
\sigma_\beta^z \rangle }\right\rvert   \frac{\hbar \omega - h +
\delta_\alpha}{\hbar \omega- h},
\end{equation}\vspace*{-5mm}
\[ \alpha,\beta = A, B, \,\alpha
\neq \beta.
\]
The quantity $\rho_{\alpha}^{(1)}(\omega)$ concerns the region $\hbar\omega
>h$, and $\rho_{\alpha}^{(2)}(\omega)$ the region $\hbar\omega<h$.

Let us consider the limits for the energies $\varepsilon_{1}(x)$ and
$\varepsilon_{2}(x)$ of the band bosonic spectrum, if their argument
changes in the interval $0\leqslant x\leqslant9$.\,\,For
definiteness, let $\delta$ be positive ($\delta>0$).\,\,The
following cases are possible.

1)~$\langle\sigma_{A}^{z}\rangle\langle\sigma_{B}^{z}\rangle>0$.

This inequality is satisfied if $h_{A}>0$ and $\,h_{B}>0$\ $(h>0)$,
or $h_{A}<0$\ and $\,h_{B}<0$\ $(h<0)$ $(h_{A}=h+\delta$,
$h_{B}=h-\delta)$.\,\,The spectral density $\rho_{\alpha}(\hbar\omega)$ differs from zero if%
\begin{equation}
h-\sqrt{\delta^{2}+9\langle\sigma_{A}^{z}\rangle\langle\sigma_{B}^{z}\rangle
t^{2}}\leqslant\hbar\omega\leqslant h-\delta
\end{equation}\vspace*{-5mm}

\noindent and\vspace*{-3mm}
\begin{equation}
h+\delta\leqslant\hbar\omega\leqslant h+\sqrt{\delta^{2}+9\langle\sigma
_{A}^{z}\rangle\langle\sigma_{B}^{z}\rangle t^{2}}.
\end{equation}
The limits of the bands are given by the maximum and minimum values
of the energies $\varepsilon_{2}(x)$ and $\varepsilon_{1}(x)$,
respectively.\,\,In the
case concerned,%
\begin{equation}
\begin{array}{l}
\min  \varepsilon_1  = \varepsilon_1(x=0) \, \equiv \,h + \delta,
          \\[2mm]
\max  \varepsilon_2  = \varepsilon_2(x=0) \, \equiv \,h - \delta.
\end{array}
\end{equation}

\noindent Those energy values determine the spectral gap (the gap
width $\Delta \varepsilon=2\delta$).\,\,The system is in the normal
phase if the chemical potential $\mu$ is located under the lower
edge of the band $\varepsilon _{2}(x)$, provided that the energies
$h_{A}$ and $h_{B}$ are positive or, if the energies $h_{A}$ and
$h_{B}$ are negative, above the upper edge of the
band $\varepsilon_{1}(x)$.\,\,The following conditions have to be satisfied:%
\[
\min\varepsilon_{2}=\varepsilon_{2}(x=9)\,\equiv\,h-\sqrt{\delta^{2}%
+9\langle\sigma_{A}^{z}\rangle\langle\sigma_{B}^{z}\rangle t^{2}}>0
\]
in the former case, and%
\[
\max\varepsilon_{1}=\varepsilon_{1}(x=9)\,\equiv\,h+\sqrt{\delta^{2}%
+9\langle\sigma_{A}^{z}\rangle\langle\sigma_{B}^{z}\rangle t^{2}}<0
\]
in the latter one (in our model, the energy of bosons is always reckoned from
the chemical potential level).

2)~$\langle\sigma_{A}^{z}\rangle\langle\sigma_{B}^{z}\rangle<0$.

At $\delta>0$, this inequality takes place if $h_{A}>0$\ and
$\,h_{B}<0$\ ($h>0$ or $h<0$).\,\,The band edges are determined now
by the
inequalities%
\begin{equation}
h-\delta\leqslant\hbar\omega\leqslant h-\sqrt{\delta^{2}-9|\langle\sigma
_{A}^{z}\rangle\langle\sigma_{B}^{z}\rangle|t^{2}}%
\end{equation}

\noindent and
\begin{equation}
h+\sqrt{\delta^{2}-9|\langle\sigma_{A}^{z}\rangle\langle\sigma_{B}^{z}%
\rangle|t^{2}}\leqslant\hbar\omega\leqslant h+\delta.
\end{equation}
The spectral gap is confined by the values%
\begin{equation}
\begin{array}{l}
\min  \varepsilon_1 = \varepsilon_1 (x = 9) =\\[3mm] h + \sqrt{\delta^2 -
9 |\langle \sigma_A^z \rangle \langle \sigma_B^z \rangle| t^2 } > 0,
                                            \\[3mm]
\max  \varepsilon_2  = \varepsilon_2 (x = 9) =\\[3mm] h - \sqrt{\delta^2
- 9 |\langle \sigma_A^z \rangle \langle \sigma_B^z \rangle| t^2 } <
0
\end{array}
\end{equation}
and the gap width equals $\Delta \varepsilon \!=\! 2 (\delta^2
\!\!-\! 9 |\langle \sigma_A^z \rangle
   \langle \sigma_B^z \rangle| \times$ $\times\, t^2 )^{1/2}$.\,\,The
chemical potential is located in the gap if the indicated
inequalities are satisfied.\,\,The gap disappears at
$\delta=\pm3t\sqrt{|\langle\sigma_{A}^{z}\rangle
\langle\sigma_{B}^{z}\rangle|}$.

The behavior of the functions $\rho_{\alpha}(\hbar\omega)$ at the
band edges is governed by both the distribution function
$\rho_{0}(x_{0})$ with the frequency-dependent argument $x_{0}$ and
the multiplier to the right from $\rho_{0}(x_{0})$ on the right-hand
side of formula (\ref{eq006}).\,\,When approaching the band edges
(including the case $x_{0}\rightarrow0$, which corresponds to the
limiting transition $\hbar\omega\rightarrow h\,\pm$ $\pm\,\delta$),
the function $\rho_{0}(x_{0})$ tends to a finite value of
$\frac{1}{\pi \sqrt{3}}$.\,\,This fact follows from formula (14),
because, in this limit, $Z_{1}(x_{0})/Z_{0}(x_{0})\rightarrow0$,
$\sqrt{Z_{0}(x_{0})}\rightarrow \frac{\sqrt{3}}{2}$,\thinspace and
$F(\pi/2,0)=\frac{\pi}{2}$.

\begin{figure}%
\vskip1mm
\includegraphics[width=7.35cm]{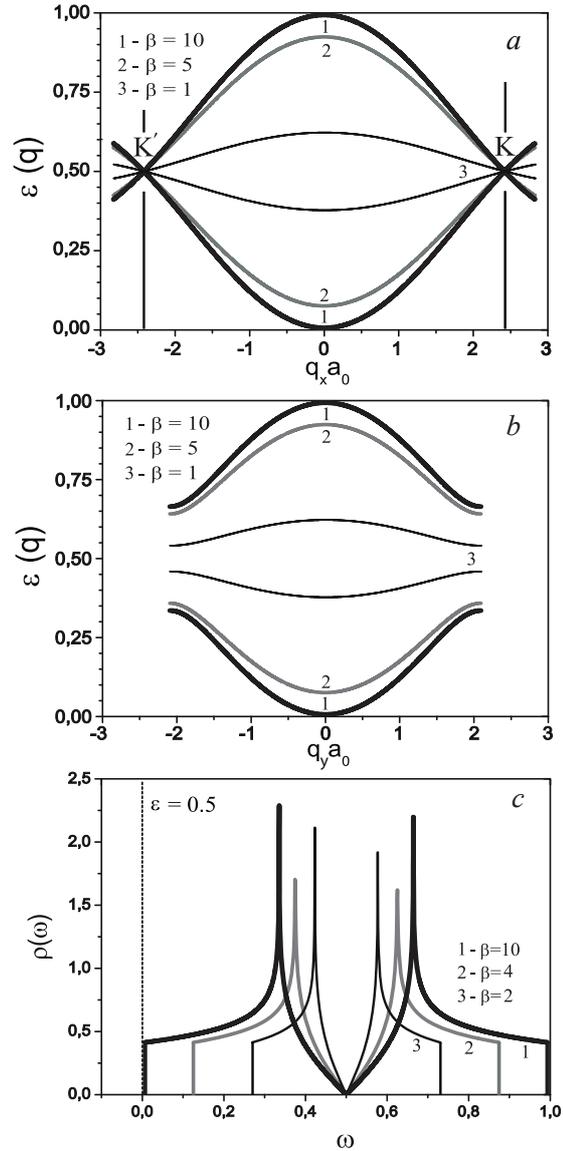}
      \vskip-3mm
      \caption{Dispersion laws of bosonic excitations in the honeycomb lattice
(\textit{a} and \textit{b}) and the frequency dependence of the
one-particle spectral density of states {$\rho(\omega)$
}(\textit{c}) in the case {$\delta=0$} and for the one-particle
energy {$\varepsilon=0.5$}.\,\,One partial spectral density of
states was obtained for the temperatures {$\beta=10, 4,$ and $2$
($\langle
n_{\alpha}\rangle=\frac{1}{2}-\langle\sigma_{\alpha}^{z}\rangle$,
$\langle n_{\alpha}\rangle_{\beta=10}=0.0066$, $\langle
n_{\alpha}\rangle_{\beta =4}=0.1192$, $\langle
n_{\alpha}\rangle_{\beta=2}=0.2689$)} }\vspace*{-2mm}
\end{figure}

On the other hand,
\[
      \hbar \omega - h + \delta_\alpha \to \left\{\!\! \begin{array}{ll}
      1, & \hbar \omega \to h + \delta_\alpha,\\[2mm]
      0, & \hbar \omega \to h - \delta_\alpha.
\end{array} \right.
\]

\begin{figure}%
\vskip1mm
\includegraphics[width=7.5cm]{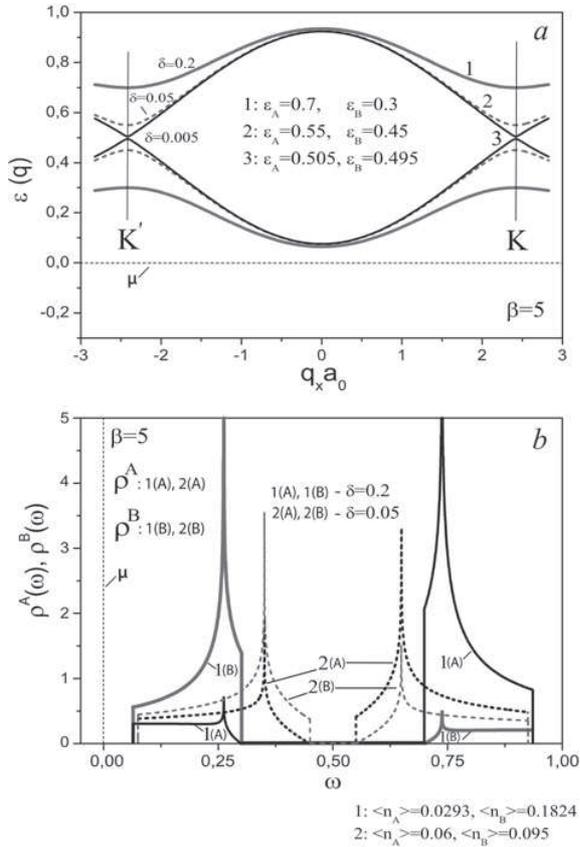}
      \vskip-3mm
      \caption{(\textit{a})~Dispersion laws
{$\varepsilon({\mathbf{q}})$} in the NO phase calculated for
{$\beta=5$ and }${\delta=0.005}${, 0.05, and 0.2}.\,\,The chemical
potential level ($\mu=0$) is located below the band spectrum (dashed
curve).\,\,(\textit{b})~Frequency dependences of the one-particle
spectral density for sublattices $A$ and $B$ calculated for
{$\beta=5$ and $\delta=0.2$ and $0.05$}}
\end{figure}
\begin{figure}%
\vskip1mm
\includegraphics[width=7.35cm]{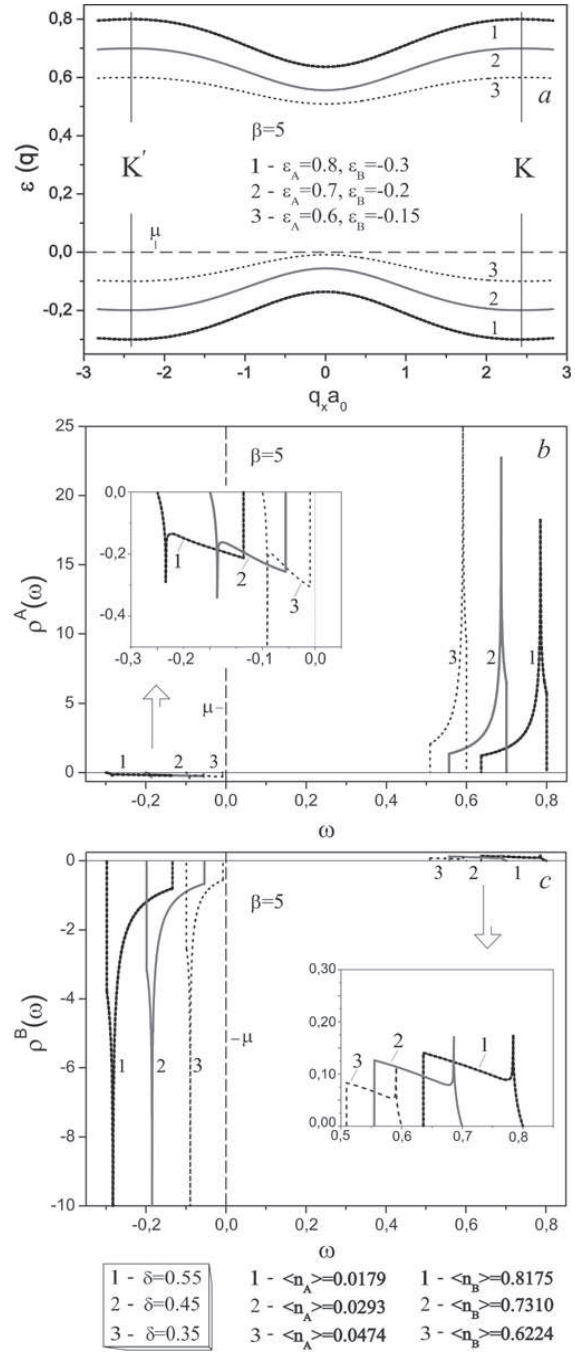}
      \vskip-3mm
      \caption{Dispersion laws for the NO phase at various
values of half-difference between one-particle energies $d=0.35$,
0.45, and 0.55 (\textit{a}), and one-particle spectral density of
states for sublattices~$A$ (\textit{b}) and
$B$ (\textit{c}) for the indicated {$\varepsilon_{A}$- and $\varepsilon_{B}$%
}-values.\,\,The average population numbers of sites in the
elementary cell, {$\langle n_{\alpha}\rangle$ ($\alpha=A,B$), are
given}.\,\,The dashed curve marks the chemical potential level
({$\mu=0$})}
\end{figure}
\begin{figure*}%
\vskip1mm
\includegraphics[width=14.6cm]{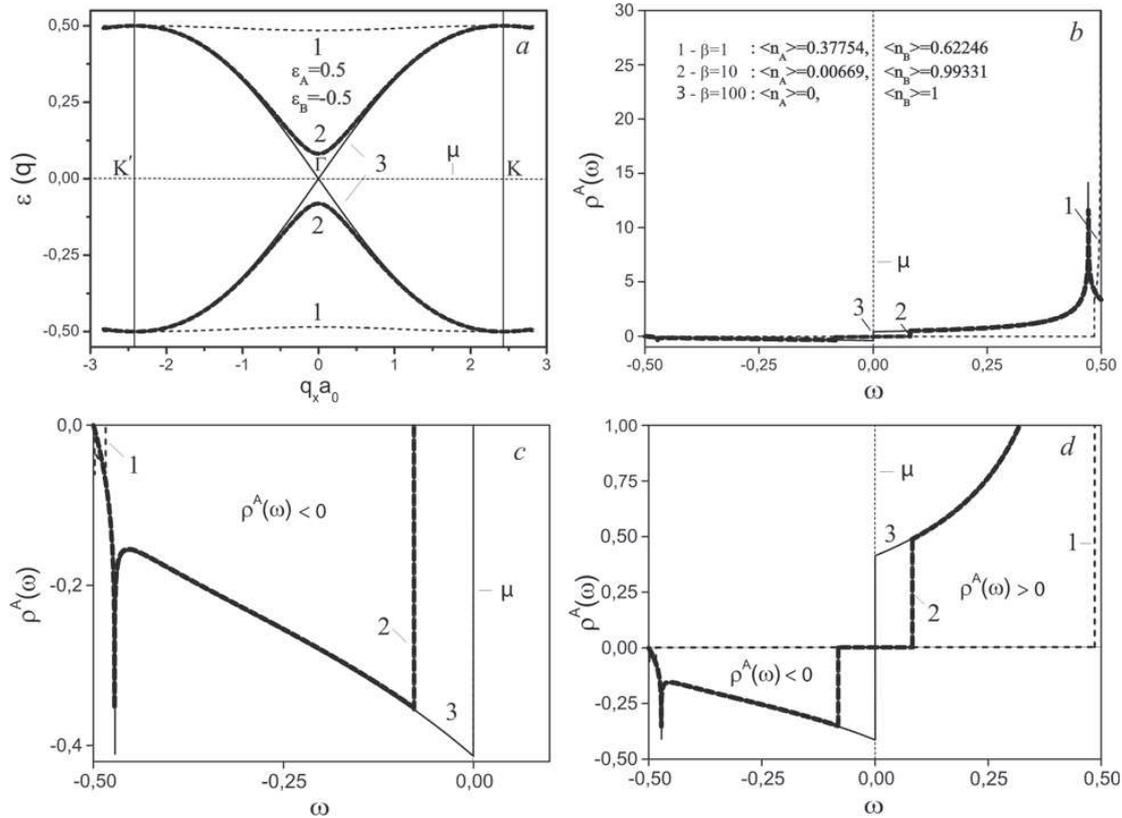}
      \vskip-3mm
\caption{(\textit{a})~Dispersion laws {$\varepsilon({\mathbf{q}})$
($\delta=0.5$)} for the temperatures {$\beta=1$, 10, and 100} (the
critical temperature {$kT_{c}=0$}) in the NO phase.
(\textit{b})~One-particle spectral density of states for sublattice
$A$, {$\rho_{A}(\omega)$.\,\,}The chemical potential level is
located at the gap middle-point.\,\,(\textit{c} and
\textit{d})~Scaled-up regions of panel \textit{b} }
\end{figure*}

\noindent Therefore,\vspace*{-3mm}
\begin{align}
\rho_{A}(\hbar\omega)  &  \rightarrow\left\{\!\!
\begin{array}
[c]{ll}%
\frac{2}{t^{2}}\frac{\delta}{|\langle\sigma_{A}^{z}\rangle\langle\sigma
_{B}^{z}\rangle|}\langle\sigma_{A}^{z}\rangle\frac{1}{\pi\sqrt{3}},
&
\hbar\omega\rightarrow h+\delta,\\
0, & \hbar\omega\rightarrow h-\delta;
\end{array}
\right. \nonumber\\[-3mm]\\[-3mm]
\rho_{B}(\hbar\omega)  &  \rightarrow\left\{\!\!
\begin{array}
[c]{ll}%
0, & \hbar\omega\rightarrow h+\delta,\\
\frac{2}{t^{2}}\frac{\delta}{|\langle\sigma_{A}^{z}\rangle\langle\sigma
_{B}^{z}\rangle|}\langle\sigma_{B}^{z}\rangle\frac{1}{\pi\sqrt{3}},
& \hbar\omega\rightarrow h-\delta.
\end{array}
\right.\nonumber
\end{align}
Expanding each of the functions $Z_{0}(x)$, $Z_{1}(x)$, and
$F(\pi/2,m)$ in a series in its argument, when the latter is small,
it can be convinced that, if the deviations from the points where
$\rho_{A,B}=0$ are small, those functions linearly increase with the
frequency.\,\,In all other cases, the function
$\rho_{A}(\hbar\omega)$ and $\rho_{B}(\hbar\omega)$ become equal to
zero at the band edges in a jump-like \mbox{manner.}\looseness=1

Numerical calculations according to formula (\ref{eq006}) and making
allowance for expressions (\ref{eq002}) for the average
$\langle\sigma_{A}^{z}\rangle$ and $\langle\sigma_{B}^{z}\rangle$
confirm the described topology of spectral densities.\,\,In
particular, in case (1) where the chemical potential is located
under or above both bands, the spectral density has a jump on one
side of the gap and grows smoothly on the other side.\,\,If the
chemical potential is located in the gap, the spectral density has
jumps on both gap sides.\,\,The general and well-known property of
the Bose--Hubbard model is that the spectral densities $\rho_{A}$
and $\rho_{B}$ are negative in the region with $\hbar\omega<0$ (i.e.
below the chemical potential level) and positive in the region with
$\hbar\omega>0$ (above $\mu$).

For a honeycomb lattice of the graphene type, the spectrum of
bosonic excitations is gapless ($\Delta\varepsilon=0$) if the depths
of potential wells are identical ($\varepsilon_{A}=$
$=\varepsilon_{B}$).\,\,Figures~3,~\textit{a} and \textit{b}
illustrate the temperature-induced variation of the spectral gap
width: as the temperature decreases, the band width increases and
reaches a maximum at the temperature of the phase transition into
the state with a Bose condensate.\,\,Two spectral branches touch
each other at Dirac points $K$ and $K^{\prime}$ in the Brillouin
zone corners.

The dispersion law of bosonic excitations for another cross-section
of the energy surface for the spectrum along the axis $q_{y}$ within
the limits of the first Brillouin zones (the component $q_{x}=0$) is
shown in Fig.~3,~\textit{b}.\,\,Here, two spectral branches do not
coincide at the Brillouin zone boundary.\,\,In the case concerned
($\delta=0$, $\varepsilon _{A}=\varepsilon_{B}=0.5$), the
one-particle spectral density for various temperatures was obtained
(Fig.~3,\textit{c}).\,\,In vicinities of the Dirac points, the
energy spectrum changes linearly (Fig.~3,~\textit{a}).

In the case of different potential well depths ($\varepsilon_{A}%
\neq$ $\neq\varepsilon_{B}$), the gap mentioned above emerges at the
Brillouin zone boun\-da\-ry.\,\,The gap width is determined by the
difference between the on-site
ener\-gi\-es.\,\,Fi\-gu\-re~4,~\textit{a} illustrates the energy
spectrum of bosonic excitations at the inverse temperature
$\beta=5$.\,\,Small differences between
the on-site energies $\delta=0.005$, 0.05, and 0.2 ($\varepsilon_{A}%
\neq\varepsilon_{B}>0$) were con\-si\-de\-red.\,\,The spectral gap
magnitude
$\Delta\varepsilon=2\delta$, and the gap limits are $\hbar\omega_{1,2}%
=h\pm\delta$.\,\,For the spectrum of bosonic excitations located
above the chemical potential level $\mu$, the calculated spectral
densities are positive (Figs.~3,~\textit{b} and 4,~\textit{b}), and,
in the case where the bands are located under the $\mu$-level, they
are ne\-ga\-ti\-ve.\,\,The limiting frequency values that confine
the interval, where $\rho_{\alpha}(\hbar\omega)\neq0,$ equal
$\hbar\omega_{3,4}=h\pm\sqrt{\delta^{2}+\langle\sigma_{A}^{z}\rangle
\langle\sigma_{B}^{z}\rangle J^{2}(0)}$.

In the case where the chemical potential level lies between the
bands (see Fig.~5,~\textit{a} corresponding to the same inverse
temperature $\beta=5$), the behavior of the energy spectrum of
bosonic excitations is essentially different.\,\,The extrema of
spectral branches at $\mathbf{q}=0$ are oriented toward the chemical
potential level $\mu$.\,\,The negative values of one-particle
spectral density ($\rho(\hbar\omega)<0$) correspond to the lower
band located under the chemical potential level $\mu$, and the
positive ones to the upper band (Figs.~5,~\textit{b} and
\textit{c}).

From the $(T,h)$ phase diagram (Fig.~1), one can see that the point
of the phase transition between the NO and SF phases, where the SF
phase becomes separated into two regions, corresponds to the
critical gap value in the spectrum of bosonic excitations,
$\Delta(kT_{c})=2\delta_{c}=1$ (in $J(0)$-units).
Figure~6,~\textit{a} (the corresponding $\delta=0.5$) illustrates
the behavior of the energy spectrum of bosonic excitations at
various temperatures in the case where the chemical potential level
is located at the band midpoint.\,\,At the inverse temperature
$\beta=100$ (practically, this is the absolute zero temperature),
two spectral branches practically touch each other at the zone
center (at $\mathbf{q}=0$); this situation corresponds to the point
of the phase transition NO$~\rightarrow~$SF for $d=0.5$ and
$\beta_{c}\rightarrow\infty$.\,\,The average population number for
Bose particles at a site in the sublattice~$A$, $\langle
n_{A}\rangle=0$, whereas in the sublattice~$B$, $\langle
n_{B}\rangle=1$.\,\,The figure also demonstrates the forms of the
one-particle spectral density of states at the $A$-site for the
values $\beta=1$ and 10, and near the critical point at
$\beta=100\,$($kT_{c}\backsimeq0$) (panels \textit{b} to
\textit{d}).

The character of changes in the frequency dependence of the
one-particle spectral density of states, which depends on the
location of the chemical potential level, qualitatively agrees with
the results of calculations obtained in the framework of the exact
diagonalization technique for the one-dimensional chain model
\cite{a17}.\,\,In the cited work, the hoppings of hard-core bosons
onto neighbor sites were considered, and negative values were
obtained for the one-particle spectral densities at energies located
below the chemical potential level.

\section{Conclusions}

On the basis of the hard-core boson model, the energy spectrum of
bosonic excitations and the one-particle spectral densities were
calculated for a plane honeycomb lattice of the graphene
type.\,\,The features in the band spectrum structure and the
spectral density in the normal (NO) phase, as well as their
dependences on the location of the chemical potential level, the
difference between the local energies of particles in the
sublattices, and the temperature, are considered.

Conditions for the appearance of a gap in the band spectrum are
analyzed.\,\,It is found that, in the case of hard-core bosons when
particles are described by the Pauli statistics, there emerges a
temperature-dependent gap (in contrast to electrons in graphene-type
lattices).\,\,The spectral gap $\Delta\varepsilon$ exists:

--~at the edge of the Brillouin zone, if the chemical potential level is
located below (above) the energy bands; in this case,
$\Delta\varepsilon =2\delta$;

--~at ${\mathbf{q}}=0,$ if the chemical potential level lies
between~~ the~~ energy~ bands; ~in this case, $\Delta\varepsilon=$ $=2\sqrt{\delta^{2}%
-|\langle\sigma_{A}^{z}\rangle\langle\sigma_{B}^{z}\rangle|J^{2}(0)}$.

 In the former case, the gap disappears at
$\delta=0$.\,\,As a result, there appear the Dirac points with a
linear dispersion law at points $K$ and $K^{\prime}$ of the
Brillouin zone.\,\,In the latter case, the gap becomes zero at $kT=$
$=0$, $h=0$, and $\delta=\frac{1}{2}J(0)$ ($\delta=\frac{1}{2}$ in
$J(0)$-units).\,\,A linear spectrum of the Dirac type, $\varepsilon_{\mathbf{q}%
}\sim\frac{J(0)}{2\sqrt{2}}aq$, also emerges in this case

The profiles of the calculated spectral densities correspond to
general criteria: the densities are negative in the interval
$\omega<0$ and positive at $\omega>0$.\,\,The specificity of the
honeycomb lattice structure manifests itself in
the available logarithmic singularities in the curves $\rho_{\alpha}%
(\hbar\omega)$ for each band and in a jump-like zeroing at the spectrum edges
(except for the points $\hbar\omega=h-\delta_{\alpha}$, where the density
tends to zero linearly).

The results of our research can serve as a basis for the description
of the thermodynamics of Bose atoms in hexagonal optical lattices
and the further study of their dynamics (experimental means that
allow the features in\ the energy spectrum and the spectral
densities of ultracold atoms in the systems of this type to be
revealed directly include the interband and momentum-resolved Bragg
spectroscopies \cite{a01,a02}).\,\,For the ultimate solution of the
problem to be obtained, it is necessary to consider the case of the
SF phase (with a Bose condensate).\,\,Unlike the normal phase, the
chemical potential in the SF phase is located in either of the
energy bands.\,\,As a result, a considerable reconstruction of the
bosonic spectrum associated with the appearance of additional
subbands occurs \cite{a18,a03,a04}.\,\,The corresponding
calculations of the dispersion laws in the bands and the spectral
densities for a lattice of the graphene type will be the subject of
our separate consideration.

\begin{figure}
\vskip1mm
\includegraphics[width=5cm]{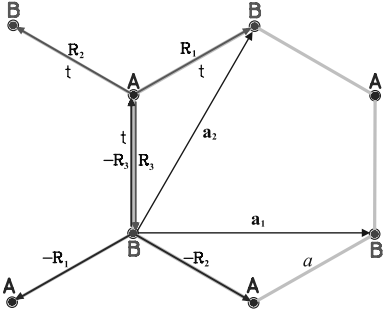}
\vskip-3mm\caption{Nearest neighbours for sites in sublattices $A$
and $B$}
    \label{figA1}
\end{figure}%
\begin{figure}
\vskip3mm
\includegraphics[width=4.5cm]{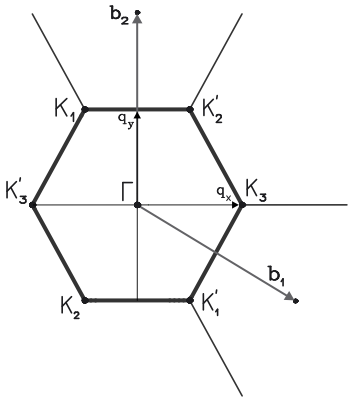}
\vskip-3mm\caption{First Brillouin zone.\,\,{${\mathbf{b}}_{1}$ and
${\mathbf{b}}_{2}$} are the translation vectors }
    \label{figA2}
\end{figure}%

\subsubsection*{\!\!\!\!\!\!APPENDIX A\\ Graphene-Type Honeycomb Lattice}

{\footnotesize The two-dimensional graphene-type honeycomb optical
lattice is obtained as a result of the interference of three
coherent laser beams \cite{a5} oriented at an angle of $2\pi/3$ with
respect to each other and with the sum of their wave vectors being
equal to zero,
$\mathbf{k}_{1}+\mathbf{k}_{2}+\mathbf{k}_{3}=0$.\,\,This lattice
include two triangular sublattices $A$ and $B$ shifted with
respect to each other by the vector $\frac{(\mathbf{a}_{1}+\mathbf{a}_{2})}%
{3}$.\,\,The elementary cell of the lattice contains two sites, one
per each sublattice.\,\,For this lattice (Fig.\,\,7), the
translation vectors are
\begin{equation}
   {\bf a_1}=(a \sqrt{3}, 0 ),~ {\bf a_2}=\left(\!a \frac{\sqrt{3}}{2},
   \frac{3}{2}a \!\right)\!\!.\tag{A1}
\end{equation}
A honeycomb lattice has a hexagonal Brillouin zone in the inverse
space of wave vectors.\,\,This is a regular hexagon with two
nonequivalent points $K$ and $K^{\prime}$ at the zone corners
(Fig.\,\,8).\,\,The corresponding translation
vectors are%
\begin{equation}
                          {\bf b_1}=\left(\!\frac{1}{\sqrt{3}a},
                            \frac{1}{3a}\!\right)\!,~ {\bf b_2}=\left(\! 0,
                            \frac{2}{3a}\!\right)\!,\tag{A2}
\end{equation}
where $\left\vert \mathbf{b_{1}}\right\vert =\left\vert \mathbf{b_{2}%
}\right\vert =\frac{2}{3a}$, and $a$ is the distance between the
neighbor sites in the direct lattice.\,\,The distance from the
Brillouin zone center to points $K$ and $K^{\prime}$ equals
$\frac{4\pi}{3\sqrt{3}a}$.

While considering the energy spectrum of quantum particles (bosons)
arranged in the optical lattice, the strong coupling approach can be
used.\,\,It is based on the consideration of particle hoppings
between the neighbor sites describing by the parameter $t$, which is
connected with the overlapping of the wave functions of Bose
particles that are localized at those sites.\,\,The coordination
number of every atom $z=3$:
\begin{equation}
          {\bf R}_1=\left(\!\frac{a\sqrt{3}}{2},
         \frac{a}{2}\!\right)\!,~
         {\bf R}_2=\left(\!-\frac{a\sqrt{3}}{2},
         \frac{a}{2}\!\right)\!,~
         {\bf R}_3=\left(0, -a \right)\!.\tag{A3}
\end{equation}
The Fourier transforms of the nearest-neighbor hopping energy
calculated in two cases~-- $A\Rightarrow B$ ($J^{AB}(\mathbf{q})$)
and $B\Rightarrow A$
($J^{BA}(\mathbf{q})$)~-- differ by the sign before the vectors $\mathbf{R}%
_{c}$ (Fig.~7):%
\begin{equation}
 J^{AB}({\bf q}) = t \sum_{c=1}^3 \mathrm{e}^{i {\bf q R}_c },~
  J^{BA}({\bf q}) = t \sum_{c=1}^3 \mathrm{e}^{-i {\bf q R}_c }.\tag{A4}
\end{equation}
Hence, we obtain%
\[
  J^{AB}({\bf q}) = t \left(\!\mathrm{e}^{-iq_y a} + 2 \cos\left(\!\textstyle \frac{\sqrt{3}}{2} q_x a \!\right)
  \mathrm{e}^{i \frac{q_y a}{2}}\!\right)
  \equiv J(\mathbf{q}),  \]
  \begin{equation}
  J^{BA}({\bf q})\equiv J(-\mathbf{q})\tag{A5}
\end{equation}
and, in the general case, the dimensionless parameter associated
with the
transfer $A\rightleftharpoons B$ between the nearest sites looks like%
\[\gamma({\bf q})= \frac{\sqrt{|J^{AB}({\bf q})\, J^{BA}({\bf
  q})|}}{t}=\]
\begin{equation}
  \label{gamma}
  =\sqrt{1 {+} 4 \cos\left(\!\textstyle\frac{\sqrt{3}}{2} q_x a\! \right)
        \cos\left(\!\textstyle\frac{3}{2} q_y a \!\right)
    {+} 4 \cos^2\left(\!\textstyle\frac{\sqrt{3}}{2} q_x a \!\right)
  }.\tag{A6}
\end{equation}
Note that $\gamma(\mathbf{q})=0$ at points $K$ and $K^{\prime}$ of
the Brillouin zone.

}

\vspace*{-5mm}
\rezume{%
І.В.\, Стасюк, І.Р.\, Дулепа, O.B.\, Величко}{ДОСЛІДЖЕННЯ
БОЗОННОГО\\ СПЕКТРА ДВОВИМІРНИХ ОПТИЧНИХ\\ ҐРАТОК ЗІ СТРУКТУРОЮ
ТИПУ\\ ГРАФЕНУ. НОРМАЛЬНА ФАЗА} {Досліджено зонний спектр
бозе-атомів у двовимірних гексагональних оптичних ґратках із
структурою типу графену. У наближенні хаотичних фаз розраховано для
нормальної фази закони дисперсії в зонах та одночастинкові
спектральні густини. Для ґратки з енергетично еквівалентними вузлами
отримано температурно залежний безщілинний спектр з точками Дірака
на краю зони Бріллюена. Хімічний потенціал розташований у цьому
випадку поза дозволеною енергетичною зоною. При відмінності між
енергіями частинок на вузлах різних підґраток, коли виникає щілина у
спектрі, хімічний потенціал може перебувати між підзонами. У такому
разі має місце значна перебудова зонного спектра. Визначено частотні
залежності одночастинкових спектральних густин для обидвох підґраток
залежно від розміщення рівня хімічного потенціалу, величини щілини у
зонному спектрі та температури.}

\end{document}